\documentclass[printer]{aa}
\usepackage{txfonts}
\usepackage{graphicx}
\usepackage{amssymb}

\def\aap{{Astronomy \& Astrophysics}}
\def\apj{{Astrophysical Journal}}
\def\apjl{{Astrophysical Journal Letters}}

\def\prd{{Phys. Rev. D}}

\def\mnras{{Monthly Notices of the Royal Astronomical Society}}
\def\nat{{Nature}}
\def\apss{{Astrophysics and Space Science}}

\titlerunning{Quark-nova remnants III}
\authorrunning{Ouyed et al.}

\begin{document}

\title{Quark-nova remnants III:\\
\small{Application to RRATs}}

\author{Rachid Ouyed$^{1}$,  Denis Leahy$^{1}$,   Brian Niebergal$^{1}$, and Youling Yue$^{1,2}$}

\institute{$^1$Department of Physics and Astronomy, University of Calgary,
2500 University Drive NW, Calgary, Alberta, T2N 1N4 Canada\\
$^2$Astronomy Department, School of Physics, Peking University, Beijing 100871, China}

\offprints{ouyed@phas.ucalgary.ca}

\date{recieved/accepted}

\abstract{
 This is the third paper of a series of papers where we explore the evolution of iron-rich
  ejecta from quark-novae.
In the first paper, we explored the case where a quark-nova ejecta forms a 
degenerate shell, supported by the star's magnetic field, with applications to SGRs.  
In the second paper  we considered 
 the case where the ejecta would have sufficient angular momentum to 
form a degenerate Keplerian torus and applied such a system
to two AXPs, namely 1E2259$+$586 and 4U0142$+$615.  
 Here we explore the late evolution of the degenerate torus and find that 
  it can remain unchanged for $\sim 10^{6}$ years before it becomes non-degenerate.
   This transition from a degenerate torus (accretion dominated) to
    a non-degenerate disk (no accretion), occurs about $10^{6}$
    years following the quark-nova, and exhibits features that are reminiscent
     of observed properties of RRATs.  
Using this model, we can account for the duration of both the radio bursts and the
quiet phase, as well as the observed radio flux from RRATs.
The unique on and off activity of the radio pulsar PSR B1931$+$24 
   is similar to that  of ``old RRATs" in our model. For old RRATs, 
 in our model, the spin-down rate during the radio loud phase
is about a factor 1.6 larger than when it is quiet, remarkably
similar to what has been measured for PSR B1931$+$24.
 We discuss a connection between XDINs and RRATs 
 and argue that some XDINs may be ``dead RRATs'' that have already consumed 
  their non-degenerate disk.  
}

\maketitle

\section{Introduction}

\cite{McLaughlin} have
recently reported the detection
of eleven ``Rotation RAdio Transients'', or ``RRATs'', characterised
by repeated, irregular radio bursts, with burst
durations of 2-30 ms, and intervals between bursts of $\sim4$~min to
$\sim3$~hr. The RRATs are concentrated at low Galactic latitudes, with
distances implied by their dispersion measures of $\sim2-7$~kpc.
For ten of the eleven RRATs discovered by \cite{McLaughlin}, an analysis of the
spacings between repeat bursts reveals an underlying spin period, $P$,
and also in three cases, a spin period derivative, $\dot{P}$. The observed
periods fall in the range 0.4 s $<P<$ 7 s, which generally overlap with
those seen for the radio pulsar population.
Since August 2003, all the sources have been reobserved at least
nine times at intervals of between one and six months (i.e. they
 show sporadic radio bursting for years). All have shown
multiple bursts, with between 4 and 229 events detected in total
from each object (see Table 1 in \cite{McLaughlin}).
For the three RRATs with values measured for both $P$ and $\dot{P}$,
a characteristic age, $\tau_c$, and a dipole surface magnetic
field, $B$, can be inferred and are listed in Table~\ref{tab_rrats}.
The total number of these objects is a few  times that
previously estimated for all radio pulsars (\cite{McLaughlin}).

         The discovery of the first X-ray counterpart
 to a RRAT was reported by \cite{rbg+06} for RRAT J1819--1458.
 The X-ray spectrum detected by Chandra and XMM-Newton is thermal, 
fit well by a blackbody with $kT \sim  0.1$ keV with a possible excess at high energies.
  Combined with
         the high inferred surface magnetic field strength, long spin period,
and lack of persistent radio emission this led to a comparison
with the population of magnetars (e.g.\cite{rbg+06}).
 These comparative studies showed the
 obvious differences between magnetars and RRATs:
 (i) the magnetar birth rate is well below
that estimated for the RRAT population (\cite{ptp06});
 (ii) X-ray properties of RRATs seem distinct from
those seen for magnetars:  RRATs seem  much colder and less luminous
than the magnetars, and
apparently lacks the hard X-ray tail seen for these
sources (e.g. \cite{2007Ap&SS.308...95G,Rea07}); (iii)
X-ray temperature of RRATs is around 0.1-0.2 keV
 below the average 0.5 keV associated with magnetars.


The lack of persistent radio emission and the long spin period of at least
 one RRAT has raised the possibility of a link between XDINs
(Haberl~2004)\footnote{No radio emission of any kind has been reported from
XDINs; recent observations (Bradley 2006) show no RRAT-like radio
bursts toward RX~J0720.4--3125 (or toward magnetars).} and RRATs.
 Furthermore, \cite{ptp06} show that the inferred birthrate
of RRATs is consistent with that of XDINs but not with magnetars.
The XDINs are slightly cooler ($kT \sim
0.04-0.1$\,keV) and less luminous ($L_X \sim 10^{31}-10^{32}$\,ergs
s$^{-1}$) than J1819--1458.  However, the measured period derivatives of two
XDINs (RBS~1223 and RX~J0720.4--3125; Kaplan \& van~Kerkwijk~2005a,
2005b), and the detection of possible proton cyclotron lines in their
spectra (van~Kerkwijk~2004), imply magnetic field strengths similar to
those of J1819--1458.  The immediate questions are (i) why
 only one RRAT  shows X-rays as the XDINs do, and (ii) why XDINs do not
show any radio emission (latest searches for pulsed and bursty
radio emission from XDINs led to no detection despite of
their proximity compared to RRATs; \cite{Kondratiev07}).


RRAT  J1819--1458 properties in the X-ray  show similarities to
  those of  radio pulsars with ages around 100~kyr.
  For example, PSR~J0538+2817 is 30~kyr old and has
$kT_{\infty} = 160$~eV, while PSR~B0656+14 is 110~kyr old and has
$kT_{\infty} = 70$~eV (see \cite{rbg+06} for details).  However,
 the inferred surface magnetic field
strength of RRATs is at least an order of magnitude
greater than the radio sources.
Two radio pulsars with comparable magnetic fields that have been
detected in X-rays are PSRs~J1718--3718 (\cite{km05}) and J1119--6127
(\cite{gkc+05}). These sources  show temperatures ($kT \sim 150-200$~eV) and
luminosities ($\sim10^{32}-10^{33}$~ergs~s$^{-1}$) comparable to that
of RRAT~J1819$-$1458, although both sources are probably much younger
(35 and 1.7~kyr, respectively) and, in contrast to RRAT~J1819--1456,
have X-ray luminosities less than their spin-down luminosities.

\subsection{Literature explanations}

The models proposed so far in the literature can be classified into 
the following categories:
 (i) Extreme pulses from distant pulsars (\cite{wsrw06}),
similar to the pulses seen from the nearby pulsar B0656$+$14.
(ii) Re-activated radio pulsars (\cite{zgd06}) where
  the RRATs are pulsars that are no longer active,
but for which a temporary ``star spot'' with multipole field components
emerges above the surface. This magnetic field component could
temporarily reactivate the radio beaming mechanism, producing the
observed bursts. 
(iii)
Nulling pulsars viewed from the opposite direction (\cite{zgd06}).
In this model, RRATs are normal pulsars with their magnetic poles
not aligned favorably for detection, but undergo 
an magnetic reversal and so occasionally 
produce emission that can be observed.
 (iv) Sporadic accretion (\cite{cs06,li06}). Here the
 RRAT mechanism might be produced by interaction of
the neutron star with an equatorial fallback disk or with orbiting
circumpolar debris.  Accretion from a disk should  quench the
radio emission mechanism, but sporadic drops in the accretion rate could
allow the radio beam to turn on for a fraction of a second, producing
the RRAT phenomenon (\cite{li06}).  

In this paper, the third of a series, we present an alternative scenario
that incorporates SGRs, AXPs, XDINs, and RRATs into one family of compact
objects.  In our model, SGRs, AXPs, XDINs, and RRATs are all strange quark stars
that differ, for the most part, only by age.

This paper is organized as follows: Sect.~2 describes our strange quark star model. 
Sect.~3  discusses the implications of the evolution of debris material 
left over from the birth of the strange quark star.
In Sect.~4 we show how after roughly one million years the debris
can become responsible for RRAT behaviour. Then, in Sect.~5 we describe
observations and make predictions using our model.
We discuss some further implications of our model in Sect.~5, 
and lastly we conclude in Sect.~6.

\section{Our model}

In the quark-nova (QN) picture 
(\cite{ouyed021, 2005ApJ...618..485K}; hereafter ODD and KOJ respectively)
the
core of a neutron star, that undergoes the phase transition to the quark phase,
shrinks in a spherically symmetric fashion to a stable, more compact strange
matter configuration faster than the overlaying material (the neutron-rich
hadronic envelope) can respond, leading to an effective core collapse.
The core of the neutron star is a few kilometers in radius initially,
but shrinks to 1-2 km in a collapse time of about 0.1 ms
(\cite{LBV}). The gravitational potential energy
released (plus latent heat of
phase transition) during this event is converted partly into 
internal energy  and partly into an outward propagating shock wave
that imparts kinetic energy to the overlying material.

As described in  a series of papers 
(\cite{OLNI,OLNII}; hereafter referred to as OLNI and OLNII),
 during a quark-nova the degenerate crust of a neutron star is blown off, 
leaving behind a strange quark star (QS) surrounded by left over, highly-metallic degenerate matter.
In OLNI we discussed the case where the ejected crust had insufficient angular momentum to escape the QS's
gravitational pull, and so would either balance with the QS's magnetic field and form 
a co-rotating shell, or fall back entirely.  Then in OLNII the case where the QS was
born with a sufficient spin-period to impart the ejected crust with enough angular
momentum  to form a degenerate torus.  In this paper, we explore the result
of this degenerate torus after enough time has passed for it to expand to densities
where it is non-degenerate.

\subsection{Emission: Vortex Expulsion vs. Accretion}

 In our model the QS is in the ground Color-Flavor Locked (CFL)
 phase and so behaves as a type II superconductor, wherein a rotationally-induced lattice
forms inside the star.
As the star spins down the magnetic field, which is confined to exist only within
the vortices, is also expelled and the subsequent magnetic reconnection leads
to the production of X-rays. The luminosity from vortex expulsion is given in OLNI to be,
\begin{equation}\label{eq:lx}
L_{\rm X, v} \simeq 2.01\times 10^{34}\ {\rm erg\ s}^{-1} \eta_{\rm X, 0.1} \dot{P}_{-11}^2 \ .
\end{equation}
Here the subscript $v$ stands for ``vortex'' and $\eta_{\rm X, 0.1}$ is the efficiency parameter,
in units of $0.1$, inherent in the conversion from magnetic
energy to observed radiation.   The spin-down rate, $\dot{P}$ is given in units of $10^{-11}$ s s$^{-1}$.

In this paper the degenerate material ejected during the quark-nova is imparted 
with sufficient angular momentum to form a degenerate torus, 
in which case accretion from this torus can result in emission that
outshines the emission due to vortex expulsion.  The condition on the initial
spin-period of the QS at birth is then an upper limit (from OLNII),
\begin{equation}\label{eq:plimit}
P_0 < P_{\rm 0, max} 
 =  2.5\ {\rm ms}\
\frac{B_{0,15}^{3/2}R_{\rm QS,10}^{9/2}}
{m_{\rm 0,-7}^{3/4}M_{\rm{QS,}1.4}^{5/4}} \ ,
\end{equation}
where $R_{\rm QS,10}$ is the radius of the QS in units of $10$ km,
$B_{0,15}$ is the initial magnetic field strength in units of $10^{15}$ G,
and the initial mass of the torus is $m_{\rm 0,-7}$ in units of $10^{-7}$ solar masses.
 From here on, we parametrize the actual period as being some fraction, $\alpha_{0.3}$,
of the maximum period in units of $0.3$; $P_0 = \alpha P_{\rm 0,max}$.

The accretion luminosity from the type of torus considered in this paper
is given by equation (13) in OLNII,
 \begin{equation}
 L_{\rm X, acc.}\simeq 1.7\times 10^{33}\ {\rm erg\ s}^{-1}
 \frac{\eta_{0.1}^{4}R_{\rm t, 15}^{6}}{\mu_{3.3}^{6}M_{\rm QS,1.4}^{4}} \ .
 \end{equation}
Here $\eta_{0.1}$ is the efficiency of conversion of accreted material into X-ray emission in units of $0.1$,
$R_{\rm t, 15}$ is the radial distance of the torus in units of $15$ km, 
$\mu_{3.3}$ is the mean molecular weight per electron in units of $3.3$ (the quiescent phase value),
and $M_{\rm QS,1.4}$ is the mass of the QS in units of $1.4$ solar masses.

\subsection{Degenerate Torus properties}

The radius of the torus can be found from the initial spin-period of the QS
as was done in OLNII,
\begin{equation}\label{eq:rtperiod}
R_{\rm t} = 15 {\rm~km~} \alpha_{0.3}^{-8/3} P_{\rm 0,ms}^{2/3} M_{\rm QS,1.4}^{1/3} ,
\end{equation}
where the initial period is in units of milliseconds.

\begin{figure}[t]
\includegraphics[angle=0,width=0.5\textwidth]{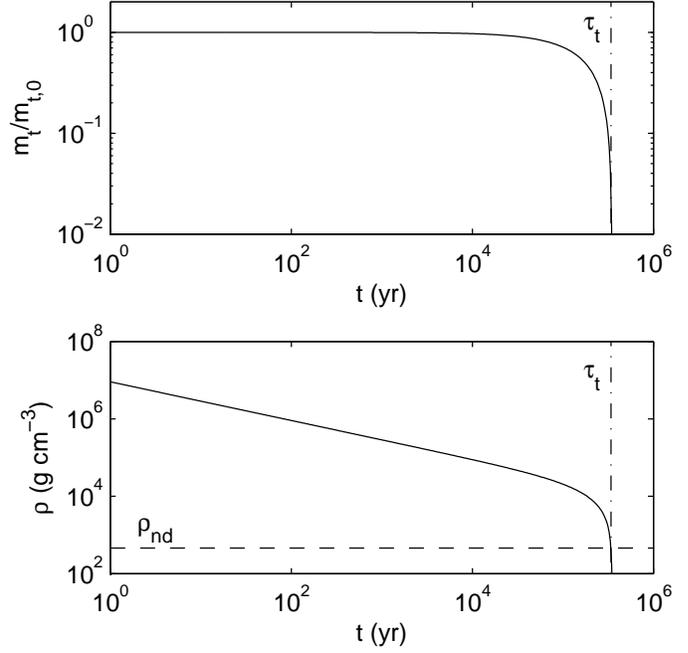}
\caption{\label{fig:mt_rho_t}
The mass and density of the torus, assuming a temperature of $1$~keV.
$\tau_{\rm t}$ and $\rho_{\rm nd}$ are the time and density at which the
torus becomes non-degenerate.
}
\end{figure}

By assuming a constant accretion rate, the evolution of the torus mass can be determined,
and the time needed for the torus to reach a density where it becomes non-degenerate is,
\begin{equation}\label{eq:tau-torus}
       \tau_{\rm t} \simeq \frac{m_0}{\dot{m}_{\rm t}}
\sim 3.4\times 10^{5}\ {\rm yrs}\ 
\frac{m_{\rm 0,-7} M_{\rm QS,1.4}^{4}\mu_{3.3}^6}
{\eta_{0.1}^3R_{\rm t,15}^6}\ ,
\end{equation}
with $\dot{m}_{\rm t}$ being the torus accretion rate given by eq(10) in OLNII.
The mass and density of the torus over time are plotted in Fig. \ref{fig:mt_rho_t}.

While the torus is above degenerate densities,
diffusion by the QS's magnetic field into the inner walls of the torus
(leading to bursting
 accretion events; see OLNII)
   increases the inner radius. 
We can estimate the change  inner radius of the torus, $R_{\rm in}= R_{\rm t}+\Delta R_{\rm t}$ (see Appendix A),  by using the typical time needed for magnetic field diffusion into the torus, $\tau_{\rm B}$,
and the typical depth into the torus at which shear stresses become great enough 
such that accretion can proceed, $\Delta r_{\rm w}$.  Both of these parameters were
estimated in OLNII (eqns. 17 \& 5).   We find that the change in inner radius over the
span of the lifetime of the torus is negligible.

While the magnetic field slowly consumes the torus' inner  edge,
 its outer edge moves outward (due to viscosity from particle collisions 
within a degenerate ideal gas) at a rate given
 by equation (A.7) in OLNII.  After the time where the torus expands to densities 
at which it becomes non-degenerate, $\tau_{\rm t}$, the torus will have extended 
 to an outer radius of,
  \begin{equation}\label{eq:Rout}
       R_{\rm out} \simeq 207\ {\rm km}\
       \frac{m_{\rm 0,-7}^{1/2} M_{\rm QS,1.4}^{2}
\mu_{\rm q,3.3}^3}{\eta_{0.1}^{3/2}R_{\rm t,15}^{7/4}}\ ,
       \end{equation}

\section{Transition to RRAT phase}
\label{sec:late}       

The evolution of the density of the torus (during its degenerate phase) 
is calculated using eq(A.8) in OLNII,
and is shown in the lower panel in Fig. \ref{fig:mt_rho_t}.
 At time $\tau_{\rm t}$, the torus density drops rapidly and eventually 
reaches densities that are below degeneracy.
For temperatures estimated in the torus (eq. 13 in ONLII;
 $T_{\rm eV}\simeq 85 \eta_{0.1}R_{\rm t,15}/M_{\rm QS,1.4}$),
the density at which the torus makes the transition from 
degeneracy to non-degeneracy is,
$\rho_{\rm nd}\simeq 11.4 \ {\rm gm~cm}^{-3}\ \eta_{0.1}^{3/2}R_{\rm t,15}^{3/2}/M_{\rm QS,1.4}^{3/2}$.

Upon the transition from degeneracy to non-degeneracy, 
the new viscosity becomes 
$\nu = 6.0\times10^{-4} T_{\rm eV}^{5/2} {\rm~cm^2~s^{-1}}$ (OLNII eq. A.5).
The corresponding accretion rate then is 
$ \dot{m} \sim  0.0057 {~\rm g~s^{-1}~} \rho_{\rm nd} H_{\rm d} T_{\rm eV}^{5/2}$
 which is completely negligible (\cite{book_acc_power});  $H_{\rm d}$ is the
 disk thickness in units of centimeters.
Thus, immediately following this transition the X-ray luminosity of
the system is no longer dominated by accretion
but rather by vortex (magnetic flux) expulsion from the star.
 The consequence of this would be a decrease in X-ray luminosity by
a few orders of magnitude in a relatively short amount of time (see left panel in Figure 2).

\subsection{Stellar properties}

The evolution of the star's period and period derivative
  are defined by the physics of vortex expulsion 
(see \cite{2006ApJ...646L..17N}),
\begin{equation}\label{eqn:pdecay}
P_{\rm{QS}}\left(t\right) = P_{\rm{QS,}0}
  \left[1 + \frac{t}{\tau_{\rm v}}\right]^{\frac{1}{3}} ,
\end{equation}
and
\begin{equation}\label{eqn:pdotdecay}
\dot{P}_{\rm{QS}}\left(t\right) = \frac{P_{\rm{QS,}0}}{3\tau_{\rm v}}
  \left[1 + \frac{t}{\tau_{\rm v}}\right]^{\frac{-2}{3}} ,
\end{equation}
where

\begin{equation}
\tau_{\rm v} =  840\ {\rm s}~
  B_{\rm QS,0,15}^{-2}
  P_{\rm QS,0,ms}^2
  M_{\rm QS,1.4}
  R_{\rm QS,10 }^{-4} ,
\end{equation}
is the characteristic age due to vortex expulsion.
The other two important relations in our model are (see OLNI and OLNII)
 \begin{eqnarray}
 B &=& \sqrt{3\kappa P\dot{P}} \nonumber \\
 P B^2 &=& P_0 B_0^2
\label{eq:B_PB2}     \ .
 \end{eqnarray}
 where $\kappa = 8.8\times 10^{33}\ {\rm G}^2 {\rm s}^{-1}$.
 The former equation describes magnetic field decay while the latter
  allows us to link the initial and current conditions.
  
By making use of the above equations and equation (\ref{eq:rtperiod}), 
the star's period, period derivative, and magnetic field can be determined 
at the time, $\tau_{\rm t}$, when the transition to a non-degenerate torus is made  (with $\tau_{\rm t} >> \tau_{\rm v}$),
   \begin{eqnarray}\label{eq:p_trans}
 P_{\rm trans.} &\simeq& 2.3\ {\rm s}\
  \frac{\mu_{\rm q,3.3}^{2} B_{0,15}^{4/3}  M_{\rm QS,1.4}^{1/2} R_{\rm QS,10}^{10/3} }
       {\eta_{0.1}R_{\rm t, 15}^{13/6}}\nonumber  \\
 \dot{P}_{\rm trans.} &\simeq& 7.3\times 10^{-14}\ {\rm s\ s}^{-1}\ P_{\rm 0, ms}
  \frac{\eta_{0.1}^{2}  R_{\rm t,15}^{13/3} }
       {\mu_{\rm q,3.3}^{4} B_{0,15}^{2/3}  M_{\rm QS,1.4}^{2} R_{\rm QS,10}^{32/3} } 
  \end{eqnarray}
  Combining equation above with eq.(\ref{eq:B_PB2}) we find
  $B_{0,15} \simeq 2.1  f_* P_{\rm trans., 3}^{35/38} \dot{P}_{\rm trans.,-13}^{13/38}$ with 
  $P_{\rm trans.}$ and $\dot{P}_{\rm trans.}$  in units
of 3 s and $10^{-13}$ s s$^{-1}$, respectively, while 
$f_* = ( R_{\rm QS,10}^{10} \mu_{\rm q,3.3} 
      M_{\rm QS,1.4}^{-1/3} \eta_{0.1}^{-6/13} \alpha_{0.3}^{8/3}  )^{-39/76}$
is of order unity. Equations (\ref{eq:rtperiod}) and (\ref{eq:p_trans}) give us
  $R_{\rm t,15}$ and $P_{\rm 0,ms}$.
With $P_{\rm 0,ms}$, $B_{0,15}$ and eq(\ref{eq:plimit}) together, we solve for  $m_{\rm 0,-7}$. 

 Table 2 shows birth parameters of the 3 RRATs studied here
 for $\alpha=0.3$. We find an initial magnetic field
 $10^{14} < B_{0}\  {\rm G} < 10^{16}$ and sub-millisecond
  birth periods. Also listed in Table 2 are the torus' initial inner
  radius, $R_{\rm t}$, and initial 
  mass, $m_{0}$ which implies $10^{-8} M_{\odot} < m_0 < 10^{-5}M_{\odot}$
   and $ 10 < R_{\rm t}\ ({\rm km}) <  35$.

\subsection{Non-degenerate disk properties}

      
  The  non-degenerate disk's thickness 
  at a radial distance $r$ is given by
$H_{\rm d} = v_{\rm th.}^{2}/g$,
where $g= GM_{\rm QS}H_{\rm d}/r^3$ is the effective gravity at
 $r$ , and
$v_{\rm th}=\sqrt{kT/(\mu_{\rm q} m_{\rm H})} \sim
 1.7\times 10^{7} {\rm cm\ s}^{-1} \mu_{\rm q,3.3}^{-1/2} T_{\rm keV}^{1/2}$,
 is the disk thermal speed. 
Thus, at $R_{\rm in}$, we find
\begin{equation}
H_{\rm d}  \sim 3.9\times 	10^4 {\rm~cm~}
\frac{ T_{\rm d, keV}^{1/2} R_{\rm in,100}^{3/2}  }
{ \mu_{\rm q,3.3}^{1/2}   M_{\rm QS,1.4}^{1/2} }\ ,
\end{equation}
 while the disk's thickness  is a few 
    hundred meters  at its outer edge, $R_{\rm out}$.    
     However, as we show below, the disk quickly cools reducing its
     temperature and thus its thickness until it solidifies.  The density would continue dropping below $\rho_{\rm nd}$ until it reaches the density of 
normal iron $\rho_{\rm Fe} \sim 7.86$ g cm$^{-3}$.
     
     The thermal evolution of the disk is given by,
\begin{equation}\label{eq:heat}
C_{\rm v}  \frac{\partial T}{\partial t}  = \frac{\Omega_{\rm d}}{4\pi}L_{\rm X,v} - L_{\rm BB,t}\ ,
\end{equation}
where $\Omega_{\rm d}=(2\pi H_{\rm d}/R_{\rm d})$ is the solid angle extended by the disk
 with thickness $H_{\rm d}$. In the above, $C_{\rm v} $ is the torus heat capacity
  while $L_{\rm BB, t}\simeq 2\pi R_{\rm out}^2 \sigma T^4$ is the torus blackbody cooling.   
 The disk  cools,  while $H_{\rm d}$ decreases, yielding smaller $\Omega_{\rm d}$
  thus further cooling.      However, the decrease in $H_{\rm d}$ stops
   when the disk gas condenses  which occurs
    at a temperature of $T_{\rm Fe, cond.}\simeq  0.265$ eV (CRC tables),
     since the disk composition is dominated by iron (see OLNI and OLNII).
     
     To find the disk's thickness at the condensation point we impose
     conservation of surface density so that
      $H_{\rm d}\rho_{\rm nd}=H_{\rm d,cond.}\rho_{\rm Fe}$;
       we assume the surface density constant during the gas phase since the gas phase
        is short-lived so that 
   \begin{equation}
H_{\rm d, cond.}  \sim 0.12  {\rm~cm~}
\frac{ R_{\rm in,100}^{3/2}  }
{ \mu_{\rm q,3.3}^{1/2}   M_{\rm QS,1.4}^{1/2} }\ .
\end{equation}   
 The solid disk is extremely thin varying from 
 a millimeter (paper thin) at $R_{\rm in}$ up to a few centimeters at its
 outer edge.  
    
Using $\Omega_{\rm d,cond.}= (2\pi H_{\rm d, cond.}/R_{\rm in})$, we estimate
   the equilibrium temperature of the condensed disk  to be
   \begin{equation}\label{eq:Td}
   T_{\rm d}\sim 18.6 \ {\rm meV} \frac{R_{\rm in,100}^{1/8}\dot{P}_{-13}^{1/2}\eta_{X,0.1}^{1/4}}{R_{\rm out, 1000}^{1/2}\mu_{\rm q,3.3}^{1/8}M_{\rm QS,1.4}^{1/8}}\ ,
   \end{equation}
   where $R_{\rm out}$ is given in units of 1000 km.
   For the 3 RRATs in the order listed listed in Tables we get, $\sim 6 $ meV,
    $\sim 86$ meV  and $\sim 5.7$  meV, respectively.
    At temperatures below 0.1 eV, condensed iron is in the form
 of ferrite, or $\alpha$-iron, a body-centered cubic structure.

\begin{table*}[hbtp]
\caption{Observational properties of the  RRATs.}
\centering
\label{tab_rrats}       
\begin{tabular}{lcc cccccccc}
\hline
&
& Star
&
&
&
&
& Radio 
&
& \\
\cline{2-4}
\cline{7-9}
Source & $P$ (s)$^1$ & $\dot{P}$ ($10^{-13}$~s s$^{-1}$)$^1$ & $L_{\rm X}$ (erg s$^{-1}$)$^2$

&&&$t_{\rm on}$ (ms)$^1$  & $t_{\rm off}$ (hr)$^1$    & $L_{\rm radio}$ (erg s$^{-1}$)$^3$
\\\hline
RRAT~J1317--5759 & 2.6 & 0.126  & $<7.5\times 10^{32}$ &&& 10&0.22 & $1.4\times 10^{31}$\\
RRAT~J1819--1458 & 4.3 & 5.76  &$ 3.3\times 10^{33}$ &&& 3 &0.057   & $5.6\times 10^{31}$\\
RRAT~J1913+3333  & 0.92& 0.0787 &$<9.4\times 10^{34}$ &&& 2 &0.21   & $2.5\times 10^{31}$\\
\hline
\end{tabular}
\\
$^1$From \cite{McLaughlin}; $^2$The observed X-ray luminosity (Reynolds et al. 2006); $^3$The observed radio luminosity assuming a bandwidth of $1$ GHz (\cite{Rea07})
\end{table*}

\begin{table*}[t!]
\caption{
Birth parameters predicted by our model (for $\alpha =0.3$).} 
\centering
\label{tab_rrats_obs_model}       
\begin{tabular}{lcc cc  ccccc}
\hline
&
& Star
&
&
&
&  Torus 
& \\
\cline{2-5}
\cline{7-8}
Source 
& $B_{0} (10^{15}$ G)
& $P_{\rm 0}$ (ms) 
& $\dot{P}_{0}$ ($10^{-6}$ s s$^{-1}$)
&
&
& $R_{\rm t}$ (km) 
& $ m_{\rm 0}\  (10^{-7}M_{\odot})$
\\
\hline
RRAT J1317--5759  &  0.91&  0.27  &  1.2 &&& 13.4  &  3.3  \\
RRAT J1819--1458  &  5.4 &  0.97 &  12 &&& 31.5   &  20   \\
RRAT J1913+3333   &  0.30&  0.20  &  0.18 &&& 10.8  &  0.53 \\
\hline
\end{tabular}

\end{table*}

\begin{table*}[t!]
\caption{
RRATs era parameters  predicted by our model (for $\alpha =0.3$).} 
\centering
\label{tab_rrats_obs_model2}       
\begin{tabular}{lcc  ccccccccccccc}
\hline
& Star
& 
&&
&
& Disk
& 
&&
& 
& Radio
&\\
\cline{2-3}
\cline{6-8}
\cline{10-13}

Source 

& $L_{\rm X,v}$ (erg s$^{-1}$)
&$T_{\rm BB}$ (eV)
&&
& $R_{\rm in}$ (km)
& $R_{\rm out}$ (km)
&$T_{\rm d}$ (meV)
&&
&$t_{\rm on}=t_{\rm acc.}$  (ms)
&$t_{\rm off}=\tau_{\rm B}$  (hr)
&$L_{\rm Radio}$ (erg s$^{-1}$)
\\
\hline
RRAT J1317--5759 & $3.19\times 10^{28}$ &  14 &&&   96 & 1082&  6.0 &&& 2.2  & 0.46     & $3.3\times 10^{31}$  \\
RRAT J1819--1458 & $6.7 \times 10^{31}$ &  100 &&&   33  & 203&  86 &&& 0.44  &0.005     & $2.3\times 10^{30}$ 
  \\
RRAT J1913+3333  & $1.25\times 10^{28}$ & 11 &&&   132 & 820&  5.7 &&& 3.5  & 1.2    &  $7.1\times 10^{31}$ 
  \\
\hline
\end{tabular}
\\
$L_{\rm X}$  from eq(\ref{eq:lx}),
$T_{\rm BB}$ from eq(\ref{eq:Tbb}),
$R_{\rm in}$ from eq(\ref{eq:Rin}),
$R_{\rm out}$ from eq(\ref{eq:Rout}),
$T_{\rm d}$ from eq(\ref{eq:Td}),
$t_{\rm acc}$ from eq(\ref{eq:tacc}),
$\tau_{\rm B}$ from eq(\ref{eq:btime}),
  $L_{\rm radio}$ from eq(\ref{eq:radioflux}) with $A_{\rm casc.}=10^2$.\\
 \end{table*}

\begin{table*}[t!]
\caption{
(Torus and RRATs) Era duration predicted by our model (for $\alpha =0.3$).} 
\centering
\label{tab_rrats_obs_model3}       
\begin{tabular}{lcc ccc }
\hline
Source 
& $\tau_{\rm torus}$ (yrs)
&$\tau_{\rm rrat}$ (yrs)
\\
\hline
RRAT J1317--5759  &   $2.2\times 10^{6}$ &   $7.9\times 10^5$\\
RRAT J1819--1458  &   $7.9\times 10^{4}$ &  $3.2\times 10^3$\\
RRAT J1913+3333   &   $1.3\times 10^{6}$ &  $4.1\times 10^5$\\
\hline
\end{tabular}
\\
$\tau_{\rm torus}$ from eq.(\ref{eq:tau-torus}); $\tau_{\rm rrat}$ from eq.(\ref{eq:trrat}).
\vspace{0.1in}
\end{table*}

\section{RRAT-like sporadic behavior}

 The inner disk will be slowly penetrated by the QS's magnetic field on
timescales determined by the induction equation,
\begin{equation}
   \frac{\partial B}{\partial t}  =  \frac{c^2}{4\pi \sigma} \nabla^2 B\ .
\end{equation}
Here, $\sigma = n_{\rm e, th} e^2 \lambda_{\rm e}/(m_{\rm e} v_{\rm rms})$
and $\lambda_{\rm e}=1/(n_{\rm e, th}\sigma_{\rm T})$, where $n_{\rm e, th}$
is the number density of thermal electrons in the disk,
$\sigma_{\rm T}$ is the Thompson scattering cross-section,
and the root-mean-square electron velocity is $v_{\rm rms}= v_{\rm th}$.
Therefore, the time needed for the magnetic field to penetrate to a depth of
 the order of  $H_{\rm d, cond.}$,

\begin{equation}\label{eq:btime}
\tau_{\rm B} \sim 1.3 \ {\rm hrs}\ 
\frac{R_{\rm in,100}^{3}}{T_{\rm d, meV}^{1/2}\mu_{\rm q,3.3}^{1/2}} \ .
\end{equation}

Once penetrated the inner disk is accreted, on free-fall timescales, 
  onto the star along the magnetic field lines with
\begin{equation}\label{eq:tacc}
t_{\rm acc} =\frac{R_{\rm in}}{v_{\rm ff}}\sim 2.3\ {\rm ms} \frac{R_{\rm in,100}^{3/2}}{M_{\rm QS,1.4}^{1/2}}\ ,
\end{equation}
where $v_{\rm ff}=\sqrt{2GM_{\rm QS}/R_{\rm in}}$ is the free-fall velocity.

\subsection{Radio emission mechanism and fluxes}

Radio emission in our model is triggered by the accretion of the inner ring
 of mass $\Delta m_{\rm d} \sim 2\pi R_{\rm in} H_{\rm d,cond.}^2 \rho_{\rm Fe}$. Each
 iron nucleon accreted leads to the generation of a GeV photon thus trigering
 a photon-pair cascade amplifying the accretion
 energy by a  factor $A_{\rm casc.}$.

The luminosity in radio is then $L_{\rm radio} = A_{\rm casc.} (0.1  \Delta m ~c^2)/t_{\rm acc.}$, or, 
\begin{equation}\label{eq:radioflux}
L_{\rm radio} \sim A_{\rm casc.}~3.7\times 10^{29}\ {\rm erg\ s}^{-1}\frac{R_{\rm in,100}^{5/2}}{\mu_{\rm q,3.3} M_{\rm QS,1.4}^{1/2}}\ .
\end{equation}
As shown in Table 3, the observed radio fluxes can be accounted for consistently
 for all 3 candidates if 
 an amplification factor of the order of $10^2$ is assumed (a
 multiplicity number for secondary pairs can be as
 high as  $10^5$; e.g. Melrose 1995 and references therein).
 We expect  the medium to be transparent to radio since it would
 most likely be emitted much above the polar region.
   We assume all radio bursts  originate at the same longitude on the
  quark star.  This guarantees that the bursts are modulated at the spin period,
   consistent with the assumption required to derive $P$
    and $\dot{P}$ from the observations of RRATs.

 Table 3 lists parameters derived from our model as compared to observations
  which shows some encouraging agreement with measured radio burst timescale
   and quiet phases. The case of RRAT J1819$-$1458 shows smaller
    bursting time and quiet time, as well as smaller radio fluxes
     probably because our model somehow predicts a smaller $R_{\rm in}$
      than the actual one. For example, using $R_{\rm in}\sim 80$ km
      instead of 33 km (i.e. $\alpha \sim 0.2$ instead of $\alpha=0.3$)
       we find much better agreement.

\section{Model Predictions}

\subsection{The two blackbodies}

One of the key aspects of
the transition from degenerate torus to a non-degenerate
disk is the shutting-off of accretion thus eliminating one
of the two BBs inherent to the accretion  era (see OLNII).
RRATs  should show 
  one dominant blackbody emission at a temperature given by
  \begin{equation}\label{eq:Tbb}
  T_{\rm BB}= (L_{\rm X,v}/(4\pi R_{\rm QS}^2\sigma ))^{1/4}\sim  0.04\ {\rm  keV} \dot{P}_{-13}^{1/2}\ ,
  \end{equation}
   where $\dot{P}$ is in units of $10^{-13}$ s s$^{-1}$,  and a dimmer one at  $T_{\rm d}$ representing the remnant iron disk.     In particular,
   for RRAT J1819-1458 for which $\dot{P}_{-13}=5.76$ we predict $T_{\rm BB}\sim 0.1$ keV
    and another blackbody in infrared at $\sim 940$ K. The blackbody temperature
    for the other two candidates are listed in Table 3 which shows a BB temperature in
     the ultra-violet region and disk temperatures in the Infrared, $50$-$100$ K.

 \subsection{Absorption lines}

Since the disk is cooler than the star ($T_{\rm d} < T_{\rm BB}$),
 it will act as an absorber of the blackbody emission.
  The column density in the inner part of the disk is
  \begin{equation}
  N_{\rm d, Fe} = \frac{H_{\rm d, cond.}\rho_{\rm Fe}}{56 m_{\rm H}}\sim 10^{22} {\rm gm\ cm}^{-2} R_{\rm in,100}^{3/2}\ ,
  \end{equation}
   high enough to lead to absorption. 
The X-ray spectrum of RRAT J1819$-$1458 showed a single
blackbody component did not fit the data because of the presence of two strong
 absorption  features around 0.5 and 1 keV (\cite{Rea07}). In our model,  these
  may  be explained as absorption lines due the disk.

\subsection{Increase in spin-down rate when radio is on}

During the quiet period the star's spin-down rate is simply
 given by $\dot{P}= P/(3 t_{\rm age})$ where $t_{\rm age}$
  is the source age. However, 
 it is natural to expect spin-down increase during the
 on era since the polar wind (secondary pairs) act as a torque against the
 star's rotation.

The angular momentum per unit mass lost at the light cylinder is $c^2/\Omega_{\rm QS}$,
which enhances the spin-down rate of the quark star during radio bursts to,
 using $(\Delta m_{\rm d}/ t_{\rm acc.}) c^2/\Omega_{\rm QS}\simeq I_{\rm QS}\dot{\Omega}_{\rm QS}$,
 $I_{\rm QS}$ is the
star's moment of inertia,
\begin{equation}\label{eq:pdoton}
 \Delta \dot{P}  \simeq 6.7\times 10^{-17}{\rm s\ s}^{-1} \frac{P^{3} R_{\rm in, 100}^{5/2}}{M_{\rm QS,1.4}R_{\rm QS, 10}^2} \ . 
\end{equation}
Compared to $\dot{P}_{\rm trans.}$ we get
\begin{equation}
\frac{\Delta \dot{P}}{\dot{P}_{\rm trans.}}\sim 9.2\times 10^{-4} \frac{P^3 R_{\rm in,100}^{5/2}B_{0,15}^{2/3}}{P_{\rm 0, ms} R_{\rm t,15}^{13/3}}
\end{equation}
 As can be seen from the equation above the increase in spin-down
  rate when radio is on is negligible is negligible. However,
  as we show below, as the RRAT ages the increase in
  spin-down rate during the on period is noticeable.
 
 \subsection{Old RRATs: {\it the case of PSR B1931$+$24}}
 
   A simple estimate of the  lifetime of a RRAT 
  can be found by integrating equation $dt/dr \simeq \tau_{\rm B}/H_{\rm d,cond.}$,
   from $R_{\rm in}$ to $R_{\rm out}$, 
  to get
\begin{eqnarray}\label{eq:trrat}
\tau_{\rm rrat}  &\sim&  
 1.6\times 10^6\ {\rm yrs} \frac{R_{\rm out, 1000}^{5/2}}{T_{\rm d,meV}^{1/2}}\ .
\end{eqnarray}
  The late stages of the RRAT  ($\tau_{rrat}\sim 10^6$ yrs) era,
   correspond to the era when the magnetic field is consuming the
    last  outermost remnants of the disk; i.e. 
  when the disk radius is close to $R_{\rm out}$.

When the radio is off (no accretion) the star spins-down
via vortex expulsion with a period derivative given by $\dot{P}= P/(3\tau_{\rm rrat})$.
Using equations (\ref{eq:trrat}) and (\ref{eq:Td})  with  $R_{\rm in}\sim R_{\rm out}$, we get 
     \begin{equation}\label{eq:pdotoff}
     \dot{P}_{\rm off}\sim 2.3\times 10^{-14}\ {\rm s\ s}^{-1} P^{4/3} R_{\rm out, 1000}^{-43/12}\ ,
     \end{equation}
     The increase in spin-down rate  during the on period,  is found by replacing $R_{\rm in}\sim
 R_{\rm out}$ in  eq.(\ref{eq:pdoton}), so that 
 \begin{equation}\label{eq:pdoton2}
 \Delta \dot{P} \simeq 2.1\times 10^{-14}{\rm s\ s}^{-1} \frac{P^{3} R_{\rm out, 1000}^{5/2}}{M_{\rm QS,1.4}R_{\rm QS, 10}^2} \ . 
\end{equation}
The spin-down rate during the on era is thus $\dot{P}_{\rm on}= \dot{P}_{\rm off}
+\Delta \dot{P}$.

The observed periodicity
of the radio-on and radio-off recurrence of  PSR B1931+24 (\cite{klo+06}) is difficult to explain
in any scenario considering an isolated pulsar, and as
\cite{klo+06} pointed out, the short shut-off time of
less than 10 seconds, rules out possible scenarios like precession
of the neutron star. Furthermore, the facts that the off periods of
PSR B1931$+$24 are five orders of magnitude longer than typical nulling periods, that the activity
pattern is quasi-periodic and that not a single null has been observed during on periods strongly
 argues against the nulling scenario.
   It appears however that  magnetospheric
conditions are sufficient to explain the change in the
neutron star torque, but it is not clear what determines the
$\sim 30$-$40$ days periodicity or what could be responsible for changing
the plasma flow in the magnetosphere, in particular in
this quasi-periodic fashion. 
     
     Assuming PSR B1931$+$24 ($P=0.813$ s) is an old RRAT that is 
      currently consuming the outermost part of its disk, 
      our model implies  a corresponding spin-down frequency during
       the off period of $\dot{\nu}_{\rm off}= - \nu^2
\dot{P}_{\rm off} \simeq  - 2.6\times 10^{-14}$ Hz s$^{-1}$, while during the on
period the increase in spin-down frequency is
  $\Delta \dot{\nu}  \simeq  - 1.7 \times 10^{-14}$ Hz s$^{-1}$.
   It means an increase of about 65\% in spin-down rate during the
    on period so that  $\dot{\nu}_{\rm on}  \simeq  - 4.3 \times 10^{-14}$ Hz s$^{-1}$.
     Our numbers are remarkably similar to what has been measured for 
  PSR B1931+24.
     For this source the observed spin-down rate the pulsar rotation slows
down 50\% faster when it is on than when it is off with 
     $\dot{\nu}_{\rm off}  =  -1.08  \times 10^{-14}$ Hz s$^{-1}$ while
       $\dot{\nu}_{\rm on}  =  -1.63 \times 10^{-14}$ Hz s$^{-1}$.

     For   PSR B1931$+$24 the quiet period last for $\sim 30$-$40$ days 
      while the active period  is of the order of $\sim 5$ days.     
 In our model,   the timescales for the bursting and quiet phases in old RRATs 
  are  $t_{\rm acc.}\sim 70$ ms and  $\tau_{\rm B}\sim 50$ days.
   While the off period is again remarkably similar to the observed
   one, the observed on period of $\sim 5$ days is difficult to reconcile with
   our model which predicts $\sim 70$ ms. It is worth noticing however
    that \cite{klo+06}
have been able to observe one switch from on to off that occurred within less than 10 seconds
 (although switches between the on and off states are rare events).
  This number is rather closer to $t_{\rm acc.}$ in our model than to the
   radio-on timescale.
  It might be possible that another mechanism is at play
   delaying the accretion onto the pole.
      Finally,  the corresponding radio flux in our model is estimated to be $L_{\rm radio}\sim
     10^{34}$ erg s$^{-1}$.

\subsection{XDINs and ``dead" RRATs in our model}

\begin{figure*}[t!]
\includegraphics[angle=90,width=.5\textwidth]{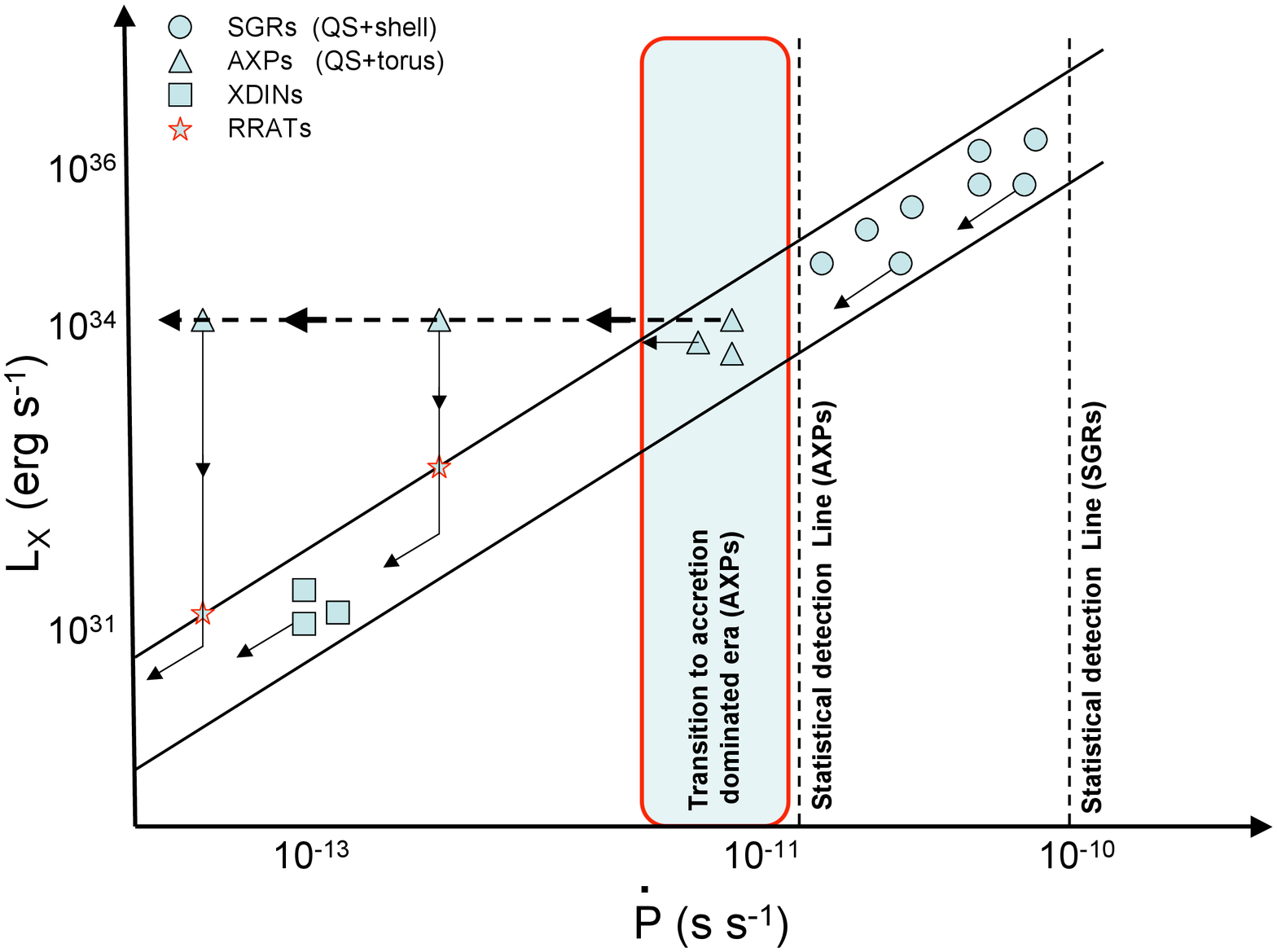}
\includegraphics[angle=0,width=.5\textwidth]{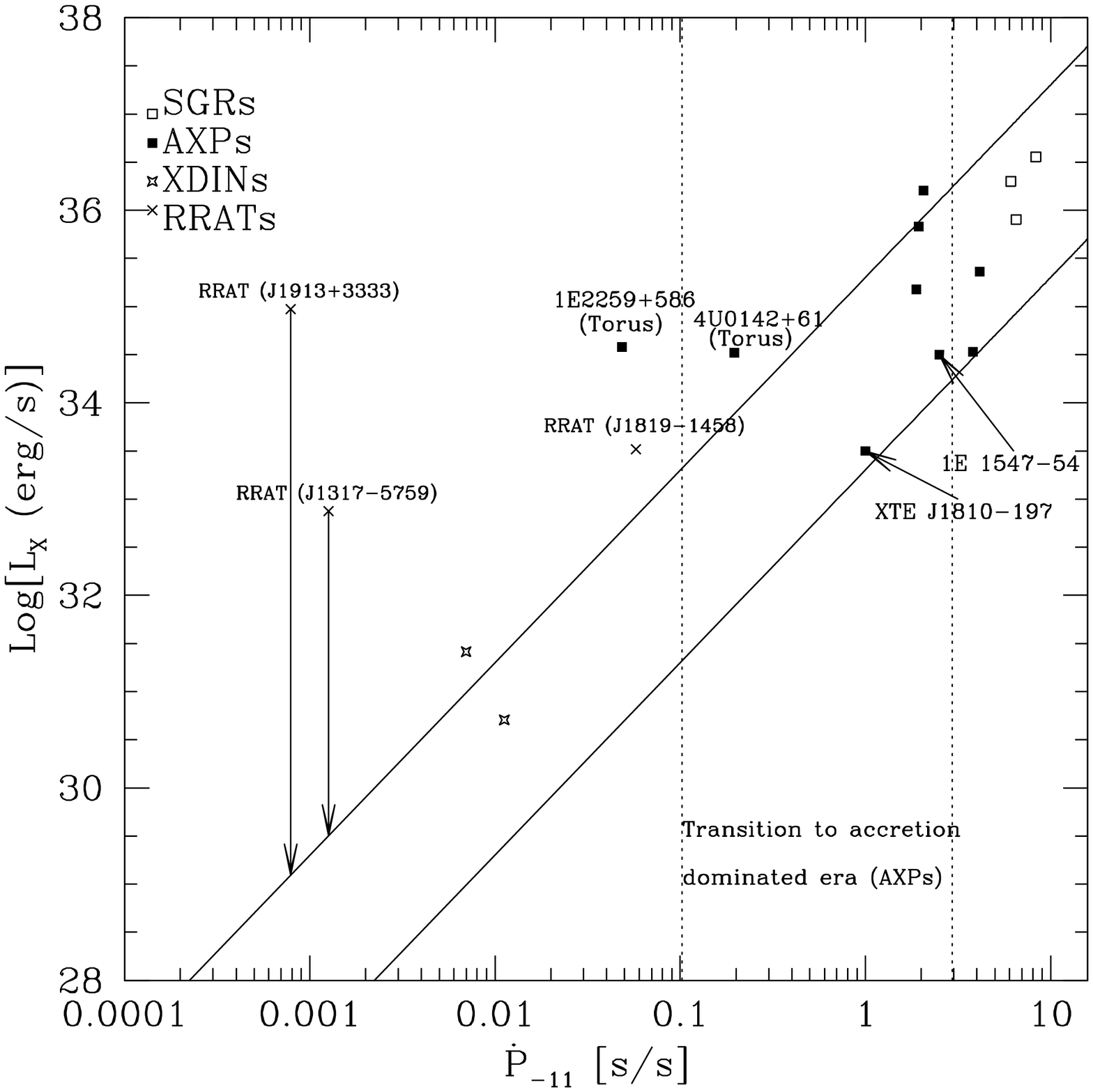}
\caption{\label{fig:illustrate}
These figures depict the important regimes of SGRs, AXPs, RRATs, and XDINs  in our model (see app. B),
the left panel being an evolutionary illustration and the right panel contains observational data.
There are two important transitions to note for AXPs
 (torus-harbouring quark stars) in our model, the first occurs when
the star spins-down to $\dot{P} \approx 10^{-11}$ s/s.  At this point
the dominant luminosity mechanism changes from being magnetic flux (vortex) expulsion
to accretion.  The second transition, occuring $\approx 10^{6}$ yrs after
the first transition, happens when the torus decreases in density enough to become
non-degenerate. At this point accretion is no longer effective, and the star's
emission is again dominated by vortex expulsion. The left-over non-dgenerate disk is 
slowly consumed resulting in RRAT-like behaviour.
Shown in the right panel are the three RRATs studied here, two of which we argue have
 spent long enough time in the tours (accretion) era before
  transiting back to having their emission dominated by vortex expulsion.
   The arrows depicts the large change in their X-ray luminosity during
    the transition.  Also, in the right panel are
the 2 knowns AXP transients (AXP 1E 1547$-$54 and AXP XTE J1810$-$197), 
which fall within the first transition, from vortex expulsion to accretion dominated emission.
 }
\end{figure*}

  Some XDINs might be direct descendants of QSs with shells (SGRs in our model;
   see Appendix B).
   These would have evolved along the ``vortex" band since their birth. However, as 
 illustrated in Figure \ref{fig:illustrate}, some XDINs might be dead RRATs. That is,
  they fell into the ``vortex" band following torus consumption and eventually
   evolved along the band.  The main difference between  SGR-descendents XDINs
    and AXP-descendent XDINs (dead RRATs) is the possibility of
     remnant disk surrounding the AXP-descendent XDINs.  
        We thus expect some of the XDINs  to share
 some common properties with RRATs such as 
 optical (or Infrared) excess.  Broad absorption lines, similar to those seen for RRAT J1819$-$1458, have been observed
for six out of seven XDINs (van Kerkwijk \& Kaplan 2007; Haberl 2007).  We have already
  argued in OLNII that these absorption lines are caused
   by absorption from  an old, cold disk (see also \S 5.2).
 
At first glance the birth rates of RRATs as descendants of AXPs in our model,
 appears to be too low to explain the high inferred population (McLaughlin et al. 2006)
  of a few times ($\sim 5$)  that of radio pulsars. Taking the RRAT birth rate to be that
   of AXPs, i.e. $\sim 1/(500\ {\rm yrs})$ (Leahy\&Ouyed 2007 and references therein), 
    implies a birth rate of $\sim 1/10$ of that of radio pulsars. In order to
     have a RRAT population $\sim 5$ times the radio pulsar population,
      the RRAT lifetime must be $\sim 50$ times longer than that
       of a radio pulsar ($\sim 10^{6}$ yrs).  In our model, the RRAT lifetime
        estimate, given by eq.(\ref{eq:trrat}), is of order a few million years. However, this
        is strongly dependent on $R_{\rm out}$. An $R_{\rm out}$ larger by a factor
         of 4 (caused by different effects; see eq.(6)) gives a large enough RRAT lifetime,
          i.e.  a low enough birth rate.

\section{Conclusion}

Here we have presented a model where RRATs fit in naturally as descendants of AXPs
with degenerate iron-rich disks, with the transition occurring when the disk becomes
non-degenerate. These AXPs in our model are the quark star remnants of quark-novae,
 surrounded by the ejected former neutron star's crust material. The evolution
 of quark stars as they spin down is summarized in the $L_{\rm X}$-$\dot{P}$ diagram
  in Figure 2 (see appendix B). Although this model is speculative, it can explain
   many features of SGRs and AXPs (OLNI and OLNII) and RRATs.


\begin{acknowledgements}
Youling Yue  thanks the University of Calgary for hosting him during this
work. This research is supported by grants from the Natural Science and
Engineering Research Council of Canada (NSERC). Youling Yue is supported by the State Scholarship Fund of China
\end{acknowledgements}


\begin{appendix}

\section{Evolution of torus inner radius}

During the degenerate phase,  the inner walls
 of the torus are carved out by magnetic penetration (leading to bursting
 accretion events; see OLNII)
   increasing the inner radius from $R_{\rm t}$ to $R_{\rm in}=
 R_{\rm t} +\Delta R_{\rm t}$ at $\tau_{\rm torus}$. 
 We can estimate $\Delta R_{\rm t}$ by integrating $dt/dr = \tau_{\rm B}/\Delta r_{\rm w}$
  up to $\tau_{\rm torus}$, with $\tau_{\rm B}$ and $\Delta r_{\rm w}$ given by 
 eq(5) and (17) in OLNII, respectively. We get

  \begin{equation}
  \frac{95}{R_{\rm t,15}^{205/24}} \simeq \left(1 + \frac{\Delta R_{\rm t, 15}}{R_{\rm t, 15}}\right)^{73/24} -1\ ,
  \end{equation}
  which leads to 2 cases
  \begin{eqnarray}\label{eq:Rin}
    \Delta R_{\rm t, 15} &\simeq& \frac{31}{R_{\rm t,15}^{181/24}}\quad\quad {\rm If}\ R_{\rm t,15} > 1.7 \\\nonumber
    \Delta R_{\rm t, 15} &\simeq& \frac{31}{R_{\rm t,15}^{132/73}} \quad\quad {\rm If}\ R_{\rm t,15} \le 1.7
  \end{eqnarray}
  
 \section{The $L_{\rm X}$-$\dot{P}$ diagram in our model}
    
\subsection{Transition from accretion dominated to RRAT era}
The left panel in Figure \ref{fig:illustrate} illustrates the
torus and RRAT eras in the $L_{\rm X}$-$\dot{P}$ diagram.
In the torus phase (lasting roughly $10^5$ yrs; eq.~\ref{eq:tau-torus}) the X-ray
luminosity is dominated by accretion from the torus, during which time the source 
spins down due to magnetic braking, thus, evolving horizontally towards smaller $\dot{P}$.  
As the torus is accreted its density decreases (due to accretion and viscous spreading), 
until it reaches non-degenercy densities, at which point accretion becomes
much less efficient, and the sources luminosity becomes 
dominated by vortex (magnetic flux) expulsion from the star.
This results in a decrease in luminosity, as is illustrated in
figure \ref{fig:illustrate} by the vertical lines.
It is this transition, we argue, that is responsible for producing RRAT behaviour.

 \subsection{Transition from vortex to accretion dominated era}
 
Before the torus phase, there exists a critical $\dot{P}$ 
at which a transition from vortex dominated to accretion dominated luminosity will occur.
This critical value is found when $L_{\rm v} \simeq L_{\rm acc}$,
\begin{equation}
\dot{P}_{\rm crit} \sim
3\times10^{-12} {\rm~s~s^{-1}~} 
\frac{\eta_{0.1}^2  R_{\rm t,15}^3   }
{\eta_{\rm X,0.1}^{1/2} M_{\rm QS,1.4}^2 },
\end{equation}  
 which implies $10^{-12} < \dot{P}_{\rm crit} ({\rm s\ s}^{-1}) < 3 \times 10^{-11}$ 
for typical values of the torus radius in our model $11 < R_{\rm t} ({\rm km})< 32$.

The two known transient AXPs (AXP 1E 1547$-$54
and AXP XTE J1810$-$197) fall within this transition
region (see right panel of fig.~\ref{fig:illustrate}). 
In an upcoming paper
we investigate transient radio-emission mechanism associated with these sources.

\subsection{Statistical appearance of SGRs}

In our model, features and statistics of SGRs can be accounted for
 if they are born (via the quark-nova mechanism) from 200 ms neutron star progenitors.
 The statistical line\footnote{The statistical line is the age of
  the youngest observed source in a population: $age\sim 1/B$ for birthrate $B$.} for 200 ms   (shell harboring) sources in
 our model  is, with a birth rate of one per $500$ yrs (as estimated by \cite{2007arXiv0710.2114L}),
\begin{eqnarray}
P_{\rm stat., 200} &\sim& 1.2\ {\rm s}\ B_{0,15}^{2/3}P_{0,{\rm 200\ ms}}^{1/3}R_{\rm QS, 10}^{4/3}M_{\rm QS,1.4}^{-1/3}\\\nonumber
\dot{P}_{\rm stat., 200} &\sim& 6\times 10^{-11} {\rm  s\ s}^{-1}\ B_{0,15}^{2/3}P_{0,{\rm 200\ ms}}^{1/3}R_{\rm QS, 10}^{4/3}M_{\rm QS,1.4}^{-1/3}\ .
\end{eqnarray}
If we adopt an initial magnetic field strength of $B_0=10^{16}$ G 
for the 200 ms objects we find a statistical line at
$P_{\rm stat., 200}\sim 5.3$ s and $\dot{P}_{\rm stat., 200}\sim 2.7\times 10^{-10}$ s s$^{-1}$,
  which is to the right of SGRs location in  the  $L_{\rm X}$-$\dot{P}$ diagram.
  In fact for these objects, as they evolve to lower $\dot{P}$
   the shell is consumed and they becomes less active.
 Thus our earlier suggestion (OLNI and OLNII) that what is referred to as AXPs in the literature
 are old shell-harbouring sources in our model that consumed most of their
 shell.

\subsection{Statistical appearance of AXPs}

The statistical line  for  millisecond (torus harboring) sources is given by eqs. (\ref{eqn:pdecay})
  and (\ref{eqn:pdotdecay}) 
 in our model (for $t\sim 500$ yrs):
 \begin{eqnarray}
P_{\rm stat., 1}&\sim& 0.2\ {\rm s}\ B_{0,15}^{2/3}P_{0,{\rm ms}}^{1/3}R_{\rm QS, 10}^{4/3}M_{\rm QS,1.4}^{-1/3}\\\nonumber
\dot{P}_{\rm stat., 1} &\sim& 10^{-11} {\rm  s\ s}^{-1}\ B_{0,15}^{2/3}P_{0,{\rm ms}}^{1/3}R_{\rm QS, 10}^{4/3}M_{\rm QS,1.4}^{-1/3}
\end{eqnarray}
However as can be seen in the right panel of Figure \ref{fig:illustrate}, all AXPs are to the right
of the above derived statistical line. As we will argue below,
 all of the observed sources to the right of the $10^{-11}$ s s$^{-1}$ line
  are SGRs (shell harboring quark stars in our model) with different
  shell masses with older ones having smaller shells thus
    with fewer and less energetic bursts (see OLNI).
  
  Torus-harbouring sources in our model are first
seen at periods of $\sim 0.2$ s and period derivative of the order of  $10^{-11}$
 s s$^{-1}$ We thus predict a population of AXPs with sub-second period
  and an X-ray luminosity $\sim 10^{34}$ erg s$^{-1}$,  from eq(\ref{eq:lx}).

    \subsection{SN transparency line}
    
     For completeness,  we also define
  a transparency line related to the time needed for a supernova remnant to be transparent to X-ray.
   A rough estimate gives
\begin{equation}
t_{\rm SN, transp.} \sim  2{\rm~yr~} M_{\rm SN, 10}v_{\rm SN,4000}^{-2}\ ,
\end{equation}
for a SN ejecta of mass $10M_{\odot}$ and velocity of $4000$ km s$^{-1}$. In
our model,  for AXPs ($P_{\rm 0, ms}\sim 1$  and $B_{0,15}\sim 1$)
 we get  $P_{\rm SN, transp.}\sim 39$ ms and $\dot{P}_{\rm SN, transp.}\sim 2.3\times 10^{-10}$ s s$^{-1}$. For SGRs ($P_{\rm 0, ms}\sim 200$  and $B_{0,15}\sim 10$),
   we find  $P_{\rm SN, transp.}\sim 1.1$ s and $\dot{P}_{\rm SN, transp.}\sim 6\times 10^{-9}$ s s$^{-1}$.  

\end{appendix}

\end{document}